\documentclass[reprint,aps,pra,superscriptaddress,nofootinbib,noeprint,longbibliography]{revtex4-2}

\usepackage{comment}
\usepackage{graphics}
\usepackage{bm}
\usepackage{color}
\usepackage{amsmath}
\usepackage{amssymb}
\usepackage{mathrsfs}
\usepackage{graphicx}
\usepackage[caption=false]{subfig}
\usepackage{enumitem}
\usepackage{hyperref} 
\usepackage{makecell} 

\usepackage{dcolumn}
\usepackage{bm}
\usepackage{multirow}
\usepackage{threeparttable}
\usepackage{hhline}

\newcommand{\TDO}{^3\Delta_1}
\newcommand{\omegaP}{\omega_\mathcal{P}}

\newcommand{\omegaPEQM}{\omega_{\mathcal{P};\text{EQM}}}
\newcommand{\Eeff}{\mathcal{E}_\text{eff}}
\newcommand{\cminv}{\text{cm}^{-1}}

\newcommand{\ThreeJ}[6]{\left(\begin{matrix}#1&#2&#3\\#4&#5&#6\end{matrix}\right)}
\newcommand{\SixJ}[6]{\left\{\begin{matrix}#1&#2&#3\\#4&#5&#6\end{matrix}\right\}}

\begin{document}

\title{Nuclear Electric Quadrupole Moment-Induced Parity Doubling \\in Molecules for Symmetry-Violation Searches}

\author{X. Fan}
\email{xingfan@g.harvard.edu}
\affiliation{Department of Physics, Harvard University, Cambridge, Massachusetts 02138, USA}
\affiliation{Harvard-MIT Center for Ultracold Atoms, Cambridge, MA 02138, USA}
\affiliation{Center for Fundamental Physics, Department of Physics and Astronomy, Northwestern University, Evanston, Illinois 60208, USA}

\author{L. Cheng}
 \email{lcheng24@jhu.edu}
 \affiliation{Department of Chemistry, Johns Hopkins University, Baltimore, MD 21218, USA}
\date{\today}

\begin{abstract}
Searches for the nuclear magnetic quadrupole moment (MQM) and nuclear Schiff moment (NSM) have high discovery potential for violations of time ($T$) and parity ($P$) reversal symmetries beyond the Standard Model.
Molecules containing heavy nuclei are typically used to enhance the sensitivity to MQMs and NSMs due to their strong internal electric fields and potential octupole deformation.
To extract these effects in the laboratory frame, a bias electric field is required to polarize the molecule by mixing states of opposite parity (parity doublets).
Typical heavy nuclei that are sensitive to symmetry-violation also possess large nuclear electric quadrupole moments (EQMs) when its nuclear spin is $I\geq1$.
We show that EQMs can significantly modify the energy splitting between parity doublet states and thus change the required polarizing electric field.
As a result, the EQM-induced energy splitting must be taken into account in designing such experiments.
We provide qualitative estimates of parity doubling from EQMs and supporting \textit{ab initio} calculations, along with implications for candidate molecules in symmetry-violation searches.
\end{abstract}

\maketitle

\section{Introduction}
\label{sec:intoduction}
Precision measurements of violations of time ($T$) and parity ($P$) reversal symmetries---referred to as symmetry violation---in atoms and molecules are powerful tools for probing physics beyond the Standard Model of particle physics (BSM)~\cite{EDMForNewPhysics2021,QuantumSensingWithMolecules2024,SearchForNewPhysicsWithAtomsAndMolecules2018,ProbingFrontiersParticlePhysicsWithTabletop-scaleExperiments2017,EDMofNucleonsNucleiAtoms2013,PTViolationInAtomicSystems2015,MoleculeQuantumScienceReview2017,GINGESFlambaum2004,neutronEDM2020,HgEDM2016}.
One prominent example is the measurement of the electron's electric dipole moment (EDM).
The most precise measurements of the electron EDM using $^{232}$ThO~\cite{ACME1Result,ACME2Result} and $^{180}$HfF$^+$~\cite{JILA1Result,JILA2Result} probe BSM physics at energy scales higher than the TeV range~\cite{InterpertingEDM2019}.
Searches for symmetry-violating effects in the nuclear sector, such as nuclear magnetic quadrupole moments (MQMs) and nuclear Schiff moments (NSMs), offer similar or even higher discovery potential~\cite{OpportunitiesForFundamentalPhysicsWwithRadioactiveMolecules2024,NSMInAtoms2002,EDMbyEnhancedNSM2020,T-SymmetryInMoleculesByMQM2014,ScreeningNSM2012,ExtensionNSM2012,EnhancedMQMinLaserCoolable2020}.
For NSM, the sensitivity is potentially enhanced by octupole-deformed nuclei such as $^{153}$Eu, $^{161}$Dy, $^{223}$Fr, $^{225}$Ra, $^{227}$Th, $^{229}$Th, and $^{229}$Pa~\cite{EnhancedMQMOctupole2022,Th229Flambaum,EDMbyEnhancedNSM2020}.
The sensitivity of a given molecule to BSM physics in spectroscopic searches is proportional to molecular sensitivity parameters, typically denoted $W_d$, $W_M$, and $W_Q$ for the electron EDM, MQM, and NSM, respectively~\cite{ParityViolationInMolecules1995,EnhancedSchiffPRC2023}.
These parameters are intrinsic to the molecular electronic structure. For example, sensitivity to the electron EDM is proportional to the effective electric field inside the molecule~\cite{HfFTimoFleig2017,HfFSkripnikov2017,HfFTitov2007,JILAEDMCandidateMolecule2006,ThOEeff84GVcm,ThOEeff75GVcm}. 
Since relativistic effects enhance molecular sensitivity to symmetry violation, molecules of interest typically contain one heavy atom bonded to an electronegative ligand~\cite{SANDARS_EDM1965,SANDARS_EDM1966}.

To extract the molecule-frame symmetry-violating effect in the laboratory frame, the molecules need to be polarized using a lab-frame electric field.
The required electric field to polarize molecules is inversely proportional to the energy splitting between two opposite-parity states~\cite{DemilleDoubletProposal2001}.
When this splitting is small, the pair of the states is called parity doublet and the splitting is called parity doubling.
In molecules, such parity doublets can arise from $\Lambda$-doublets, $\Omega$-doublets~\cite{PbOInvestigation2000,JILAEDMCandidateMolecule2006,ThOProposal_2011}, $\ell$-doublets~\cite{PolyEDMProposal_2017}, or $K$-doublets~\cite{RaOCH3RotVib2022,RaOCH3Phelan2021}.
The $\Lambda$- and $\Omega$-doublets arise from nearly degenerate electronic states with opposite electronic angular momentum~\cite{Brown_Carrington_2003}, while the $\ell$- and $K$-doublets arise from vibrational and rotational states with opposite vibrational or rotational angular momentum, respectively~\cite{HirotaTextbook2012}.
In the two most recent successful electron EDM searches using $^{232}$ThO~\cite{ACMEProposal_2010,JILAEDMCandidateMolecule2006,ACME1Result,ACME2Result} and $^{180}$HfF$^+$~\cite{JILA1Result,JILA2Result}, $\Omega$-doublet in the ground $\TDO~(\Omega=\pm1)$ states were used.
The parity doubling is 360~kHz for $^{232}$ThO and 740~kHz for $^{180}$HfF$^+$. 
Combined with their lab-frame Stark shift coefficient of $d_\text{lab}\simeq1$~MHz/(V/cm), an electric field of 10--100~V/cm is sufficient to fully polarize them~\cite{ACMETechnicalPaper2017,JILASystematic2023}, more than two orders of magnitude smaller than those used in the EDM experiments without a parity doublet~\cite{YbF2011Result,YbFMethodAndAnalysis2012, TlEDMResult1,TlEDMResult2,TlEDMResult3}.
Using smaller electric fields has been critical for (i) simplifying the experimental setup~\cite{ACMETechnicalPaper2017}, (ii) minimizing the electric field-correlated shift of the magnetic field from leakage and charging currents from electrode bias~\cite{TlEDMResult3}, (iii) suppressing systematic errors from geometric phase and the $\mathbf{v}\times\mathbf{E}$ term~\cite{BerryPhaseCommins1991}, (iv) reducing the electric field-induced $g$-factor, and (v) enabling the use of ion traps for EDM searches~\cite{JILASystematic2023,JILA_Proposal2011,JILADemonstration2013}.
All of these factors are essential for the current state-of-the-art electron EDM searches and will become even more important for future EDM, MQM, and NSM searches as precision increases.
In particular, the ability to use ion traps---which can achieve high sensitivity even with a small number of ions---may be crucial for experiments using rare and radioactive octupole-deformed nuclei.

In this paper, we show that the nuclear electric quadrupole moment (EQM) of the heavy nuclei could modify the parity doubling in molecules, thereby changing the electric field required for polarization.
Isotopes with $I=0$ have been used for the electron EDM searches (e.g. $^{232}$Th and $^{180}$Hf), but $I\geq1/2$ is required for NSM searches and $I\geq1$ is required for MQM searches.
Many heavy nuclei relevant for MQM and NSM searches also have large nuclear EQMs~\cite{EnhancedMQMOctupole2022,EQMTable2021}.
Since the EQM strongly couples to the electronic wavefunction, it modifies the parity doubling in $\Lambda$-doublets and $\Omega$-doublets in combination with spin-orbit coupling.
This affects symmetry-violation searches using $\Omega$-doublet structures, such as $^{177}$HfF$^+$, $^{179}$HfF$^+$, $^{181}$TaN, $^{181}$TaO$^+$, $^{229}$ThO, and $^{229}$ThF$^+$~\cite{EDM_QLS_Yan_2024,Th229Flambaum,MQMDeformedNuclei2018}.
The requirement of the larger electric field could be particularly relevant in molecular ions, where the maximum applicable electric field is lower than in neutral systems~\cite{JILASystematic2023}.
Ref.~\cite{HfFCPViolationCalculation2018} examined the EQM-induced parity doublings in $^{177}$HfF$^+$ and $^{179}$HfF$^+$ using estimates based on measured $g$-factor and atomic quadrupole coupling matrix elements; here we complement the study in Ref.~\cite{HfFCPViolationCalculation2018} with \textit{ab initio} calculations of the nuclear quadrupole-coupling parameters.
We also investigate EQM-induced parity doubling in the $\tilde{X}^2\Sigma(010)$ state of polyatomic molecules, such as $^{173}$YbOH~\cite{PolyEDMProposal_2017,YbOHCharacterizing_2023,YbOHMQM_TheoreticalStudy2019,EnhancedMQMinLaserCoolable2020}, $^{223}$RaOH~\cite{Th229Flambaum}, and $^{175}$LuOH$^+$~\cite{LuOH2022Proposal,LuOHMQMEstimate2020}.
In these polyatomic molecules, the EQM-induced parity doubling arises in combination with the vibronic coupling (the so-called Renner-Teller effect).
We first provide qualitative estimates of this effect, then perform \textit{ab initio} computation, and finally discuss its implications for symmetry-violation searches.
\section{Theory and Computation}
\label{sec:Calculation}
In the following, we take the electron EDM as an example, but the argument applies to MQM and NSM searches.
In a simple two-state model using the parity eigenbasis $|\pm\rangle$, the Hamiltonian for a parity doublet is given by
\begin{equation}
    H=\begin{pmatrix}
\omegaP/2 & -d_\text{lab}\mathcal{E}_\text{lab}-d_e\Eeff\\
-d_\text{lab}\mathcal{E}_\text{lab}-d_e\mathcal{E}_\text{eff} & -\omegaP/2
\end{pmatrix},
\end{equation}
where $\omegaP$ is the parity-doubling parameter, $d_\text{lab}$ is the lab-frame electric dipole moment, $\mathcal{E}_\text{lab}$ is the applied lab-frame electric field, $d_e$ is the symmetry-violating effect of interest, and $\Eeff$ is the effective electric field in the molecule.
The eigenvalues are
\begin{eqnarray}    
    E_\pm&=&\pm\frac{\omegaP}{2}\sqrt{1+\left(\frac{d_\text{lab}\mathcal{E}_\text{lab}+d_e\mathcal{E}_\text{eff}}{\omegaP/2}\right)^2}.
\end{eqnarray}
The full EDM sensitivity is achieved when $d_\text{lab}\mathcal{E}_\text{lab}\gg\omegaP/2$, where $E_\pm\simeq \pm(d_\text{lab}\mathcal{E}_\text{lab}+d_e\mathcal{E}_\text{eff})$.
The characteristic electric field required for this regime is
\begin{equation}
    \mathcal{E}_\mathcal{P}\equiv\frac{\omegaP}{2d_\text{lab}}.\label{eq:Ep}
\end{equation}
For example, $70$~\% of the full EDM sensitivity is achieved when $\mathcal{E}_\text{lab}=\mathcal{E}_\mathcal{P}$.

The magnitude of the parity doubling $\omegaP$ is determined by the coupling between the two states that form the parity doublet.
For instance, in the $\Omega$-doublet, the states $|\Omega=\pm1\rangle$ form the parity doublet.
Suppose the parity doublets are formed by\footnote{Hereafter, we assume a specific sign convention for simplicity, but the discussion in this manuscript is not affected by the sign of the linear superposition.}
\begin{subequations}    
    \begin{align}        
|+\rangle&=\frac{1}{\sqrt{2}}\left(|\Omega=+1\rangle + |\Omega=-1\rangle\right)\\
    |-\rangle&=\frac{1}{\sqrt{2}}\left(|\Omega=+1\rangle - |\Omega=-1\rangle\right),
    \end{align}
\end{subequations}
without a coupling between $|\Omega=+1\rangle$ and $|\Omega=-1\rangle$, the parity doublet states $|\pm\rangle$ are perfectly degenerate and $\omegaP=0$.
If a coupling term $H'$ couples $|\Omega=\pm1\rangle$ with
\begin{equation}
    |\langle\Omega=\pm1|H'|\Omega=\mp1\rangle|\equiv W_\mathcal{P},
\end{equation}
the resulting parity doubling is
\begin{equation}
    \omegaP=|\langle+|H'|+\rangle-\langle-|H'|-\rangle|=2W_\mathcal{P}
\end{equation}
Thus, to achieve a small $\omegaP$ and a small polarizing electric field [Eq.~\eqref{eq:Ep}], the coupling between the two states must be weak.
In the following, we provide estimates for two important examples: the $\Omega$-doublet in the $\TDO$ state and the $\ell$-doublet in the $\tilde{X}^2\Sigma(010)$ state in polyatomic molecules.

\subsection{$\Omega$-doublet}
\label{sec:OmegaDoublet}
The two most sensitive electron EDM measurements were performed in the $\TDO$ state of $^{232}$ThO and $^{180}$HfF$^+$\cite{ACME2Result,JILA2Result}.
The parity doublet is formed by $|\pm\rangle = \frac{|\Omega = +1\rangle \pm |\Omega = -1\rangle}{\sqrt{2}}$.  
The $\TDO$ state follows Hund’s case (a), and we label molecular states using the quantum numbers $|\Sigma, \Lambda, \Omega\rangle$.
Here, $\Sigma$ ($\Lambda$) is the projection of the electron spin (orbital) angular momentum onto the molecular axis $\hat{n}$, and $\Omega = \Sigma + \Lambda$~\cite{Brown_Carrington_2003}. 
In this basis, the $\Omega = \pm1$ states correspond to:
\begin{equation}
|\Omega=\pm1\rangle=|\Sigma=\mp1,\Lambda=\pm2,\Omega=\pm1\rangle.
\end{equation}
Coupling between $|\Omega = \pm1\rangle$ breaks the degeneracy of the parity doublet states $|\pm\rangle$~\cite{OmegaDoublingAndLimit2014}.  
Even in the absence of hyperfine coupling, degeneracy breaking comes from two $L$-uncoupling terms ($H_\text{LU} = -2B\,\boldsymbol{J} \cdot \boldsymbol{L}$) and two spin-orbit coupling terms ($H_\text{SO} = A\,\boldsymbol{L} \cdot \boldsymbol{S}$).
The $L$-uncoupling (LU) and spin-orbit (SO) coupling terms induce state mixing with the following selection rules: $\Delta\Sigma=0,~\Delta\Lambda=\pm1,~\Delta\Omega=\pm1$ for $H_\text{LU}$ and $\Delta\Sigma=\pm1,~\Delta\Lambda=\mp1,\Delta\Omega=0$ for $H_\text{SO}$, respectively. 
An example of coupling between $|\Omega=\pm1\rangle$ arises from a fourth-order perturbation
\begin{align}    
    \frac{1}{\Delta\omega_e^3}&\langle-1,+2,+1|H_\text{SO}|0,+1,+1\rangle\nonumber\\
    &\times\langle0,+1,+1|H_\text{LU}|0,0,0\rangle\nonumber\\
    &\times\langle0,0,0|H_\text{LU}|0,-1,-1\rangle\nonumber\\
    &\times\langle0,-1,-1|H_\text{SO}|+1,-2,-1\rangle,
\end{align} 
where $\Delta \omega_e$ is the typical energy difference between electronic states.
We denote this path as
\begin{align}    
|\Sigma,\Lambda,&\Omega\rangle=|-1,+2,+1\rangle\xrightarrow{\text{SO}}|0,+1,+1\rangle\nonumber\\
&\xrightarrow{\text{LU}}|0,0,0\rangle\xrightarrow{\text{LU}}|0,-1,-1\rangle\xrightarrow{\text{SO}}|+1,-2,-1\rangle.\nonumber\\\label{eq:NormalDoublingPath_OmegaDoublet}
\end{align} 
An example path that corresponds to Eq.~\eqref{eq:NormalDoublingPath_OmegaDoublet} using the molecular term symbols is
\begin{equation}    
\TDO\xrightarrow{\text{SO}}{^1\Pi_1}\xrightarrow{\text{LU}}{^1\Sigma}\xrightarrow{\text{LU}}{^1\Pi_1}\xrightarrow{\text{SO}}{\TDO}.
\end{equation} 
There are 12 possible paths from the order $H_\text{SO}$ and $H_\text{LU}$ couplings ($_4C_2=6$) and whether each $H_\text{SO}$ changes the spin ($\Delta S=0$ or $\pm1$), such as
$\TDO\xrightarrow{\text{LU}}{^3\Pi_0}\xrightarrow{\text{SO}}{^1\Sigma}\xrightarrow{\text{LU}}{^1\Pi_1}\xrightarrow{\text{SO}}{\TDO}$, $\TDO\xrightarrow{\text{SO}}{^3\Pi_1}\xrightarrow{\text{LU}}{^3\Sigma}\xrightarrow{\text{LU}}{^3\Pi_1}\xrightarrow{\text{SO}}{\TDO}$, and so forth\cite{cossel2014Thesis}.
They can either add constructively or cancel each other, but the resulting parity doubling is approximately
\begin{equation}
    \omegaP=2W_\mathcal{P}\simeq 2
\times12\frac{B^2A^2}{\Delta\omega_e^3} J(J+1)
\end{equation}
as an order-of-magnitude estimate.
Using typical parameters for EDM-sensitive molecules, $B \simeq 0.3~\cminv$, $A \simeq 500~\cminv$, $\Delta\omega_e \simeq 5000~\cminv$, and $J = 1$, we estimate $\omega_\mathcal{P}/2\pi\simeq1.7\times10^{-5}~\text{cm}^{-1}=260$~kHz, which is in agreement with the actual $\omegaP$ in $^{232}$ThO ($\omega_\mathcal{P}/2\pi=360$~kHz)~\cite{ACMETechnicalPaper2017} and $^{180}$HfF$^+$ ($\omega_\mathcal{P}/2\pi=740$~kHz)~\cite{JILASystematic2023}.

The situation changes when the heavy nucleus has a nuclear spin $I\geq1$ and thus an EQM.
The EQM Hamiltonian is
\begin{align}
    H_\text{EQM}=-eT^2(\boldsymbol{\nabla E})\cdot T^2(\boldsymbol{Q}),
\end{align}
where $T^2(\boldsymbol{\nabla E})$ is the rank-2 electric field gradient  and $T^2(\boldsymbol{Q})$ is the rank-2 nuclear quadrupole tensor.
The perpendicular component of the electric field gradient,  
$T^2_{\pm2}(\boldsymbol{\nabla E})$, changes the electronic angular momentum by two quanta:  \begin{equation}
    \Delta\Lambda=\pm2~~\text{from}~T^2_{\pm2}(\boldsymbol{\nabla E}).
\end{equation}
This introduces additional coupling between the $\Omega = \pm 1$ states in $\TDO$~\cite{Brown_Carrington_2003}.
Two mechanisms contribute:
\begin{subequations}
    \begin{align}
    |\Sigma,\Lambda,\Omega\rangle&=|-1,+2,+1\rangle\xrightarrow{\text{SO}}|0,+1,+1\rangle\nonumber\\
    &\xrightarrow{\text{EQM}}|0,-1,-1\rangle\xrightarrow{\text{SO}}|+1,-2,-1\rangle.\label{eq:EQM_Omega_a}\\
    |\Sigma,\Lambda,\Omega\rangle&=|-1,+2,+1\rangle\xrightarrow{\text{SO}}|0,+1,+1\rangle\nonumber\\
    &\xrightarrow{\text{SO}}|+1,0,+1\rangle\xrightarrow{\text{EQM}}|+1,-2,-1\rangle.\label{eq:EQM_Omega_b}    
    \end{align}
\end{subequations} 
The first mechanism [Eq.~\eqref{eq:EQM_Omega_a}] involves spin-orbit admixture of the $|0,\pm 1,\pm 1\rangle$ states into the $|\mp1,\pm 2,\pm 1\rangle$ states and EQM-induced mixing between $|0,+1,+1\rangle$ and $|0,-1,-1\rangle$.  
In the second mechanism [Eq.~\eqref{eq:EQM_Omega_b}], $|\Sigma, \Lambda, \Omega\rangle = |-1,+2,+1\rangle$ obtains the admixture of  
$|+1,0,+1\rangle$ via two spin-orbit coupling terms, and the EQM interaction couples between $|+1,0,+1\rangle$ and $|+1,-2,-1\rangle$.  
(The same mechanism generates another path with opposite signs: $|+1,-2,-1\rangle\xrightarrow{\text{SO}}|0,-1,-1\rangle\xrightarrow{\text{SO}}|-1,0,-1\rangle\xrightarrow{\text{EQM}}|-1,+2,+1\rangle$.)
In the molecular spectroscopic notation, the corresponding paths for Eqs.~\eqref{eq:EQM_Omega_a} and \eqref{eq:EQM_Omega_b} are
\begin{subequations}
    \begin{align}
    &\TDO\xrightarrow{\text{SO}}{^1\Pi_1}\xrightarrow{\text{EQM}}{^1\Pi_1}\xrightarrow{\text{SO}}{\TDO}
    \end{align}  
and
\begin{align}
    &\TDO\xrightarrow{\text{SO}}{^1\Pi_1}\xrightarrow{\text{SO}}{^3\Pi_1}\xrightarrow{\text{EQM}}{\TDO}
    \end{align}
\end{subequations} 
correspondingly.
Note that the contributions from these two mechanisms can either add constructively or cancel each other, requiring {\it{ab initio}} calculations (Sec.~\ref{sec:Abinitio}).  

The resulting EQM-induced parity splitting $\omegaPEQM$ is estimated as
\begin{align}    
    \omegaPEQM&\simeq\frac{4A_1A_2}{\Delta\omega_{e;1}\Delta\omega_{e;2}} \left|\langle\eta|{H}_\mathrm{EQM}|\eta'\rangle\right|\times\mathcal{O}(1)\label{eq:omegaPOmegaDoubletEstimate}.
\end{align}
Here, $\eta$ and $\eta'$ are the intermediate states; $\eta=|0,+1,+1\rangle$ and $\eta'=|0,-1,-1\rangle$ in Eq.~\eqref{eq:EQM_Omega_a}, and $\eta=|+1,0,+1\rangle$ and $\eta'=|+1,-2,-1\rangle$ in Eq.~\eqref{eq:EQM_Omega_b}.
$A_{i}$ and $\Delta\omega_{e;i}~(i=1,2)$ are the spin-orbit coupling and energy difference in the perturbation paths, $\langle\eta|{H}_\mathrm{EQM}|\eta'\rangle$ is the EQM coupling, and the $\mathcal{O}(1)$ factor contains information about the alignment of the nuclear spin with the internuclear axis, calculated in Sec.~\ref{sec:Mixing}.

For the molecules considered here, $\langle\eta|{H}_\mathrm{EQM}|\eta'\rangle\simeq0.005\text{--}0.03~\cminv$, $A\simeq500~\text{cm}^{-1}$, and $\Delta\omega_e\simeq5000~\text{cm}^{-1}$, giving an estimate of
\begin{equation}
    \frac{\omegaPEQM}{2\pi}\simeq 5\text{--}30\mathrm{~MHz}.
\end{equation}
The EQM effect is thus potentially significant for $\Omega$-doublets in molecules for symmetry violation searches.
The estimate presented here will be quantified using {\it{ab initio}} calculations in Sec.~\ref{sec:Abinitio}.

\subsection{$\ell$-doublet}
\label{sec:lDoublet}
In linear polyatomic molecules such as YbOH, RaOH, and LuOH$^{+}$, the parity doublet arises from the vibrational angular momentum, known as the $\ell$-doublet.
Typically, the parity doublet in the vibrationally excited state $(v_1,v_2^\ell,v_3) = (01^{\pm1}0)$ of the electronic ground state $^2\Sigma$ is used for EDM measurements \cite{PolyEDMProposal_2017}.
Here, $v_1$ and $v_3$ are the vibrational quantum numbers for the two stretching modes, $v_2$ corresponds to the bending mode, and $\ell$ is the projection of the vibrational angular momentum onto the molecular axis.

The ground $^2\Sigma$ states of these molecules are typically well described by Hund’s case (b), and we label the states as $|\Sigma;\ell,\Lambda,K\rangle$, where $K \equiv \Lambda + \ell$, and $\Sigma=\pm1/2$ in the molecules here.
The parity doublets correspond to  
\begin{subequations}
    \begin{align}        
|\pm\rangle=\frac{1}{\sqrt{2}}\bigg(|&\Sigma;\ell=+1,\Lambda=0,K=+1\rangle\nonumber\\
&\pm|\Sigma;\ell=-1,\Lambda=0,K=-1\rangle\bigg).
    \end{align}
\end{subequations}  
The EQM effect can induce additional parity doubling together with a vibronic interaction, the so-called ``Renner–Teller" (RT) coupling: $H_\text{RT}=V_{11}(Q_+L_-+Q_-L_+)$~\cite{HirotaTextbook2012}, where $Q_\pm$ are raising and lowering operators for the vibrational angular momentum.
The selection rules for RT coupling are $\Delta \ell = \mp1, \Delta\Lambda = \pm1$, and $\Delta K = 0$.  
This interaction leads to a third-order coupling between the $|\Sigma;\ell=\pm1,\Lambda=0,K=\pm1\rangle$ states via two possible paths.
For $\Sigma=+1/2$, for example,
\begin{subequations}
\begin{align} 
    |\Sigma;\ell,\Lambda,K&\rangle=|+\tfrac{1}{2};-1,0,-1\rangle\xrightarrow{\text{RT}}|+\tfrac{1}{2};0,-1,-1\rangle\nonumber\\
    &\xrightarrow{\text{EQM}}|+\tfrac{1}{2};0,+1,+1\rangle\xrightarrow{\text{RT}}|+\tfrac{1}{2};+1,0,+1\rangle\label{eq:EQM_Omega_a_poly}\\
    |\Sigma;\ell,\Lambda,K&\rangle=|+\tfrac{1}{2};-1,0,-1\rangle\xrightarrow{\text{RT}}|+\tfrac{1}{2};0,-1,-1\rangle\nonumber\\
    &\xrightarrow{\text{RT}}|+\tfrac{1}{2};+1,-2,-1\rangle\xrightarrow{\text{EQM}}|+\tfrac{1}{2};+1,0,+1\rangle.\label{eq:EQM_Omega_b_poly}
    \end{align}
\end{subequations} 
In the molecular spectroscopic notation, the corresponding paths for Eqs.~\eqref{eq:EQM_Omega_a_poly} and \eqref{eq:EQM_Omega_b_poly} are
\begin{subequations}
    \begin{align}
    {^2\Sigma(010)}\xrightarrow{\text{RT}}&~{^2\Pi_{1/2}(000)}\nonumber\\
    &~\xrightarrow{\text{EQM}}{^2\Pi_{3/2}(000)}\xrightarrow{\text{RT}}{^2\Sigma(010)}
    \end{align}
and
\begin{align}     {^2\Sigma(010)}\xrightarrow{\text{RT}}&~{^2\Pi_{1/2}(000)}\nonumber\\
    &~\xrightarrow{\text{RT}}{^2\Delta_{3/2}(010)}\xrightarrow{\text{EQM}}{^2\Sigma(010)}
    \end{align}
\end{subequations} 
correspondingly.
As in (Sec.~\ref{sec:OmegaDoublet}), the contributions of these two paths depend on the energy levels and specific coupling between the states and are calculated quantitatively in Sec.~\ref{sec:Abinitio}.
The RT coupling parameter $V_{11}$ is generally smaller than or comparable to the spin-orbit coupling constant $A$.
For example, in the coupling between $\tilde{X}^2\Sigma(010)$ and $\tilde{A}^2\Pi(000)$,  
$V_{11} \simeq 200~\text{cm}^{-1}$ in RaOH and YbOH, and $V_{11} \simeq 1500~\text{cm}^{-1}$ in LuOH$^+$, leading to generally smaller EQM-induced doubling in these polyatomic molecules.

The resulting EQM-induced parity splitting is estimated in a similar way as Eq.~\eqref{eq:omegaPOmegaDoubletEstimate}.
\begin{align}    
\omegaPEQM&\simeq\frac{4V_{11;1}V_{11;2}}{\Delta\omega_{e;1}\Delta\omega_{e;2}}\left|\langle \eta|H_\text{EQM}|\eta'\rangle\right|\times\mathcal{O}(1). \label{PEQML}
\end{align}
Here, $\eta$ and $\eta'$ are the intermediate states; $\eta = |+\tfrac{1}{2};0,-1,-1\rangle$ and $\eta' = |+\tfrac{1}{2};0,+1,+1\rangle$ in Eq.~\eqref{eq:EQM_Omega_a_poly}, and $\eta = |+\tfrac{1}{2};+1,-2,-1\rangle$ and $\eta' = |+\tfrac{1}{2};+1,0,+1\rangle$ in Eq.~\eqref{eq:EQM_Omega_b_poly}.
$V_{11;i}$ and $\Delta\omega_{e;i}~(i=1,2)$ are the Renner-Teller coupling and energy difference in the perturbation paths, $\langle\eta|{H}_\mathrm{EQM}|\eta'\rangle$ is the EQM coupling, and the $\mathcal{O}(1)$ factor contains information about the alignment of the nuclear spin with the internuclear axis, calculated in Sec.~\ref{sec:Mixing}.

For the molecules considered here, $\langle \eta|H_\text{EQM}|\eta'\rangle\simeq 0.005\text{--}0.03~\cminv$, $\Delta\omega_e \simeq 15000~\text{cm}^{-1}$, and $V_{11} \simeq 200~\text{cm}^{-1}$, giving an estimate of  
\begin{equation}
    \frac{\omegaPEQM}{2\pi}\simeq 0.1\text{--}0.6\mathrm{~MHz}
\end{equation}
for $^{173}$YbOH and $^{223}$RaOH, and two orders of magnitude larger for $^{175}$LuOH$^+$ due to its larger RT coupling $V_{11}\simeq1500$~cm$^{-1}$.
These estimates will again be confirmed via {\it{ab initio}} calculations in Sec.~\ref{sec:Abinitio}.

\subsection{\textit{Ab initio} calculations}
\label{sec:Abinitio}

 All \textit{ab initio} calculations presented here have been performed using the CFOUR program package \cite{CFOURfull,Matthews20a}. To investigate the contributions from nuclear electric quadrupole coupling to $\Omega$-doublets of the $\TDO$ states, we perform calculations for the $\Omega=+1$ and $\Omega=-1$ components of the $\TDO$ states in HfF$^+$, ThO, ThF$^+$, TaO$^+$, and TaN.
Relativistic effects have been taken into account using the exact two-component Hamiltonian with atomic mean-field integrals (the X2CAMF scheme)\cite{Liu18,Zhang22}. 
X2CAMF calculations incorporate spin-orbit coupling in the molecular orbitals.
Namely, they provide variational treatments of spin-orbit coupling in the spinor representation.
The wave functions for the $|^3\Delta_1, \Omega=\pm 1 \rangle$ states 
are dominated by
the $|S=1, \Sigma=\mp 1,\Lambda=\pm 2,\Omega=\pm 1\rangle$ components and include the admixture of the other spin components
with $\Omega=\pm 1$. 
In other words, the computed $|^3\Delta_1, \Omega=\pm 1 \rangle$ wave functions 
include
the contributions from all spin components with $\Omega=\pm 1$ 
\begin{eqnarray}
    |^3\Delta_1, \Omega=\pm 1 \rangle
    &=& c_0 |1, \mp 1,\pm 2,\pm 1\rangle + c_1 |1, 0,\pm 1,\pm 1\rangle \nonumber \\
    &+& c_2 |0, 0,\pm 1,\pm 1\rangle + c_3 |1, \pm 1, 0, \pm 1\rangle \nonumber \\
    &+& \cdots\label{eq:Omega_Mixture_All},
\end{eqnarray}
with $c_0$ close to 1 and $c_n$'s taking small values.
Therefore, the EQM-induced coupling computed here
\begin{align}    
    q_2=-2\sqrt{6}\langle\TDO,\Omega=\pm 1 |T^2_{\pm2}(\nabla\mathbf{E})|\TDO,\Omega=\mp 1 \rangle    \label{eq:q2_abinitio_Omega}
\end{align}
includes the contributions from both mechanisms described in Eqs.~\eqref{eq:EQM_Omega_a} and \eqref{eq:EQM_Omega_b}.

We have used X2CAMF equation-of-motion coupled-cluster singles and doubles (EOM-CCSD) \cite{Stanton93a,Asthana19} method to calculate the $|^3\Delta_1, \Omega=\pm 1 \rangle$ wave functions, taking the closed-shell $^1\Sigma^+$ state wave function as the reference state.
The $|^3\Delta_1, \Omega=1 \rangle$ state is then obtained by exciting a $\sigma_{1/2}$ spinor to a $\delta_{3/2}$ spinor. Similarly, the $|^3\Delta_1, \Omega=-1 \rangle$ state is obtained by exciting a $\sigma_{-1/2}$ spinor to a $\delta_{-3/2}$ spinor. In this way, both $|^3\Delta_1, \Omega=\pm 1 \rangle$ wave functions are represented using the same set of molecular spinors. 
This simplifies the evaluation of transition properties between these two states. 
Relativistic EOM-CCSD calculations have been shown to provide accurate $q_2$ values for heavy-atom-containing molecules \cite{Cheng15}.
However, we note that the transition matrix elements between the dominant spin components $|S, \Sigma,\Lambda,\Omega\rangle=|1, \mp 1,\pm 2,\pm 1\rangle$ do not contribute to the $q_2$ values targeted here.
Therefore, the values of $\langle ^3\Delta_1, \Omega=\pm 1 |eQq_2|^3\Delta_1, \Omega=\mp 1 \rangle$ entirely depends on small admixture of the other spin components.
The X2CAMF-EOM-CCSD wave functions provide reasonably accurate description of these admixtures.
For example, the X2CAMF-EOM-CCSD $g$-factors for HfF$^+$, ThF$^+$, and ThO amount to 0.0097, 0.056, and 0.0035, respectively, in reasonable agreement with measured values of 0.0118 \cite{JILADemonstration2013}, 0.048 \cite{ThF+Spectroscopy2022}, and 0.0088 \cite{ThOHStateCharacterization2011,Kirilov13}. 
Thus, it is safe to use the computed values as order-of-magnitude estimates.

The $eQq_2$ value of 42~MHz for $^{177}$HfF$^+$ computed here is around a factor of 1.8 smaller than the value of 78~MHz in Ref.~\cite{HfFCPViolationCalculation2018}, which is obtained by multiplying the value of 110~MHz reported in Ref.~\cite{HfFCPViolationCalculation2018} by a factor of $\sqrt{1/2}$ to account for the difference in the definitions of $eQq_2$.
We obtained a similar value of 75 MHz when adopting the same assumption as in Ref. \cite{HfFCPViolationCalculation2018} that the corrections to the $g$-factor of the $^3\Delta_1$ state is entirely attributed to the admixture of $^3\Pi_1$ state, i.e., multiplying an EOM-CCSD $eQq_2$ value of 5378 MHz for the $^3\Pi_1$ state with a factor of 0.0118+0.0023=0.0141.
On the other hand, the other spin components have smaller $eQq_2$ values. Their admixtures to the $^3\Delta_1$ wave function contribute less to the $eQq_2$ value. Therefore, we view the $eQq_2$ value in \cite{HfFCPViolationCalculation2018} as an upper bound for this parameter. 


We obtain the EQM matrix elements between the $\ell$-doublet states in linear triatomic molecules including YbOH, RaOH, and LuOH$^+$ by performing
calculations of the linear vibronic coupling constants and electronic EQM matrix elements involved in Eqs. \eqref{eq:EQM_Omega_a_poly} and \eqref{eq:EQM_Omega_b_poly}. 
Specifically, we perform EOM-CCSD calculations for the linear vibronic coupling constants \cite{Ichino09} between the $\tilde{X}^2\Sigma^+$ and $\tilde{A}^2\Pi$ states and between the $\tilde{A}^2\Pi$ the $^2\Delta$ states [$V_{11}$'s in Eq. (\ref{PEQML})] and for the $eQq_2$ values of the $\tilde{A}^2\Pi$ states and the $eQq$ coupling matrix elements between $\tilde{X}^2\Sigma^+$ and $^2\Delta$ states [$q_2(\eta,\eta')$'s in Eq. (\ref{PEQML})].

Using the notation $|\ell,\Lambda,K\rangle$, the computed wavefunctions for the vibrational first excited states of the bending mode within the ground electronic state, $|\tilde{X}^2\Sigma^+(010)\rangle$, include the contributions from a variety of vibronic wave functions with $K=\pm 1$
\begin{eqnarray}
    |\tilde{X}^2\Sigma^+(010),K=\pm1\rangle
    &=& c_0 |\pm 1,0,\pm 1\rangle\nonumber\\
    &+& c_1 |0,\pm 1,\pm 1\rangle \nonumber \\
    &+& c_2 |1, \mp 1, \pm2, \pm 1\rangle \nonumber \\
    &+& \cdots\label{eq:Ell_Mixture_All},
\end{eqnarray}
in which $c_0$ is close to 1, $c_1$ represents a first-order admixture of $|^2\Pi, \ell=0, \Lambda=\pm 1\rangle$ through RT coupling, and $c_2$ is a coefficient for the second-order admixture of a $|^2\Delta, \ell=\mp1, \Lambda=\pm 2\rangle$ state. The EQM-induced coupling is then computed as
\begin{align}    
    &q_2=-2\sqrt{6}\nonumber\\
    &\times\left\langle^2\Sigma^+(010),\ell=\pm1|T^2_{\pm2}(\nabla\mathbf{E})|^2\Sigma^+(010),\ell=\mp1\right\rangle.\label{eq:q2_abinitio_Ell}
\end{align}
The spin-free exact two-component theory in its one-electron variant (the SFX2C-1e scheme) \cite{Dyall01,Liu09,Cheng11b} has been used to account for scalar-relativistic effects in these calculations.
SFX2C-1e-EOM-CCSD calculations have been demonstrated to provide accurate linear vibronic coupling constants between the $\tilde{B}^2\Sigma^+$ state and the $\tilde{A}^2\Pi$ states in CaOH, SrOH, and YbOH\cite{ChengDoylePolyCalculation2021}.
We expect the computed coupling constants between the ground $\tilde{X}^2\Sigma^+$ state and the $\tilde{A}^2\Pi$ states in the molecules studied here to be accurate. 
Based on the benchmark SFX2C-1e-EOM-CCSD calculations of $eQq_2$ values of $^2\Pi$ states containing heavy elements\cite{Cheng15}, 
we take a relative error of 50\% as a conservative error estimate for the computed EQM matrix elements. We emphasize again that we use the computed values as order-of-magnitude estimates for the EQM matrix elements in the present analysis.   



\subsection{Hyperfine Structure}
\label{sec:Mixing}
The resulting parity splitting $\omegaPEQM$ depends on the alignment of the heavy nucleus's spin $\boldsymbol{I}$ (e.g., Th in ThO and Hf in HfF$^+$) with the molecular axis (i.e. hyperfine structure).
For the Hund's case (a) $\Omega$-doublets, $\boldsymbol{\Omega}$ couples to the nuclei's rotational angular momentum $\boldsymbol{R}$ to form the total rotational angular momentum $\boldsymbol{J}=\boldsymbol{\Omega}+\boldsymbol{R}$.
$\boldsymbol{J}$ then couples to the nuclear spin $\boldsymbol{I}$ to form $\boldsymbol{F_1} = \boldsymbol{J} + \boldsymbol{I}$.  
When the other nucleus also has a nuclear spin $\boldsymbol{I_2}$ (e.g., F in HfF$^+$), $\boldsymbol{F_1}$ and $\boldsymbol{I_2}$ combine to form the total angular momentum $\boldsymbol{F}$.
The EQM-induced effect from the heavy nucleus does not depend on $\boldsymbol{I_2}$, $\boldsymbol{F}$ and $m_F$ (projection of $\boldsymbol{F}$ in the lab frame), so we omit these quantum numbers in the following.

The coupling between $\boldsymbol{I}$ and $\boldsymbol{J}$ from magnetic hyperfine interaction (e.g. $\boldsymbol{I}\cdot\boldsymbol{J}$) or axial electrical quadrupole moment $eQq_0$ mixes different $J$ states.
The eigenstate including these interactions can be labeled by its dominant rotational angular momentum $\tilde{J}$, and it is a superposition of different $J$ states, $|\Omega,\tilde{J},F_1\rangle\equiv\sum_Ja^{\tilde{J},F_1}_J|\Omega,J,F_1\rangle$.
For the lowest rotational state, the admixture of $J=2$ into the $\tilde{J}=1$ state is $(A_{||}/4B)^2$.
For the considered molecules, $A_{||}\lesssim10$~GHz and $B\gtrsim7.5$~GHz, so the admixture of $J=2$ is less than $10~\%$.
The $eQq_0$-induced mixing can also be estimated similarly to be small.
Thus, to estimate $\omegaPEQM$ for energy eigenstates, we approximate $|\Omega,\tilde{J},F_1\rangle\simeq|\Omega,J,F_1\rangle$ and focus on its dominant rotational state.

\begin{figure}
    \centering
\includegraphics[width=\linewidth]{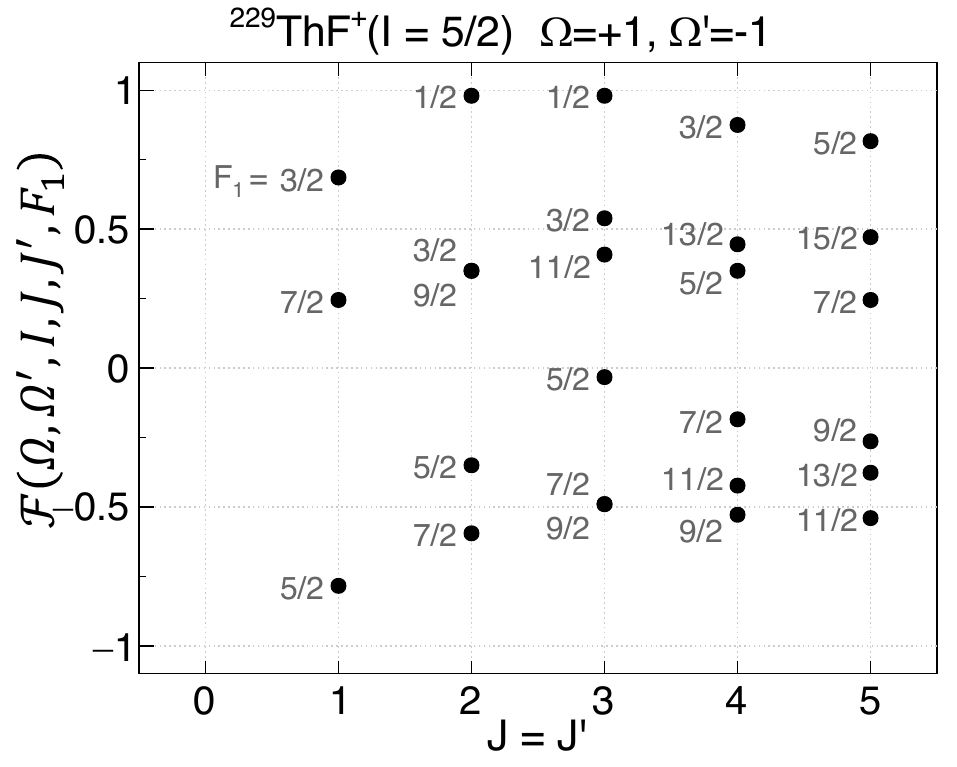} 
     \caption{The nuclear spin–molecular axis alignment factor $\mathcal{F}(\Omega,\Omega',I,J,J',F_1)$ [Eq.~\eqref{eq:MatrixElementF}] for the $\TDO$ state in $^{229}$ThF$^{+}$ ($I=5/2$). We set $J=J'$ as an example to show its magnitude. The numbers next to the dots show $F_1$.}
    \label{fig:F_function}
\end{figure}
Since $H_\text{SO}$ does not change $\Omega$, the EQM-induced splitting $\omegaPEQM$ can be calculated from the simple expectation value of $H_\text{EQM}$ for the energy eigenstate $|\TDO,\Omega,J,F_1\rangle$ [Eq.~\eqref{eq:Omega_Mixture_All}].
Using the $q_2$ calculated in Eq.~\eqref{eq:q2_abinitio_Omega}, $\omegaPEQM$ is \cite{VibrationalCalculation2022,Brown_Carrington_2003}
\begin{align}    
    \omegaPEQM=2\bigg|\langle\TDO,&\Omega=+1,J,F_1|\nonumber\\
    &\times H_\text{EQM}|\TDO,\Omega'=- 1,J,F_1\rangle\bigg|\nonumber\\
    =\bigg|\frac{2eQq_2}{4\sqrt{6}}&\times\mathcal{F}(\Omega=+1,\Omega'=-1,I,J,J,F_1)\bigg|,
\end{align}
where
\begin{align}    
    \mathcal{F}&(\Omega,\Omega',I,J,J',F_1)\equiv(-1)^{J'+I+F_1+J-\Omega}\SixJ{I}{J'}{F_1}{J}{I}{2}\nonumber\\
    &\times\ThreeJ{I}{2}{I}{-I}{0}{I}^{-1}\sum_{q=\pm2}\ThreeJ{J}{2}{J'}{-\Omega}{q}{\Omega'}\sqrt{(2J+1)(2J'+1)}.\label{eq:MatrixElementF}
\end{align}
$\mathcal{F}(\Omega,\Omega',I,J,J',F_1)$ contains information about the alignment of the nuclear spin with the molecular axis.
Fig.~\ref{fig:F_function} shows the calculated $\mathcal{F}(\Omega,\Omega',I,J,J',F_1)$ for $^{229}$ThF$^+$ ($I=5/2$) with $J'=J$.
The alignment factor is $\mathcal{O}(1)$ depending on $J$, $I$, and $F_1$.

Similarly, the alignment factor for the $\tilde{X}^2\Sigma(010)$ $\ell$-doublet states can also be calculated.
The nuclear spin $\boldsymbol{I}$ of the heavy nucleus couples to the electron spin $\boldsymbol{S}$, forming $\boldsymbol{G} = \boldsymbol{I} + \boldsymbol{S}$.
This, in turn, couples to the rotational angular momentum $\boldsymbol{N} = \boldsymbol{\Lambda} + \boldsymbol{\ell} + \boldsymbol{R}(=\boldsymbol{K} + \boldsymbol{R})$, forming $\boldsymbol{F_1}=\boldsymbol{N}+\boldsymbol{G}$.
Finally, $\boldsymbol{F_1}$ couples to the nuclear spin of the other nucleus, $\boldsymbol{I_2}$ (e.g. H in YbOH, RaOH, and LuOH$^+$) to form the total angular momentum $\boldsymbol{F}$.
Again, the EQM-induced effect from the heavy nucleus does not depend on $\boldsymbol{I_2}$, $\boldsymbol{F}$ and $m_F$, so we omit these quantum numbers in the following.

\begin{figure}
    \centering
\includegraphics[width=\linewidth]{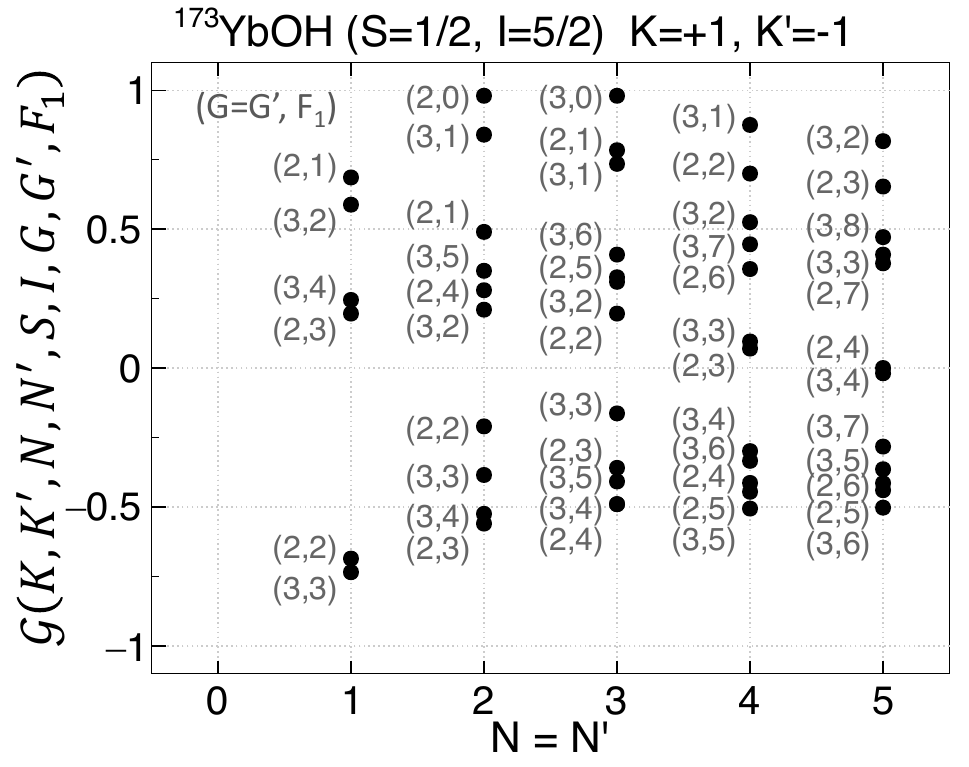} 
     \caption{The nuclear spin–molecular axis alignment factor $\mathcal{G}(K,K',N,N',S,I,G,G',F_1)$ [Eq.~\eqref{eq:MatrixElementG}] for the $\tilde{X}^2\Sigma(010)$ state in $^{173}$YbOH ($S=1/2$ and $I=5/2$). We set $N=N'$ and $G=G'$ as an example to show its magnitude. The numbers next to the dots show ($G=G'$, $F_1$).}
    \label{fig:G_function}
\end{figure}
Following the same discussion in the $\Omega$-doublet's case, we drop the mixture of other angular momentum ($N$ and $G$) and calculate $\omegaPEQM$ for specific $N$ and $G$.
Since $H_\text{RT}$ does not change $K$, the EQM-induced parity splitting $\omegaPEQM$ can be calculated from the simple expectation value of $H_\text{EQM}$ for the energy eigenstate $|^2\Sigma^+(010),K,N,S,I,G,F_1\rangle$ [Eq.~\eqref{eq:Ell_Mixture_All}].
Using the $q_2$ calculated in Eq.~\eqref{eq:q2_abinitio_Ell}, $\omegaPEQM$ is \cite{YuikiMagicField2023,Brown_Carrington_2003}
\begin{align}    
    \omegaPEQM~~~&\nonumber\\
    =2\bigg|\langle^2\Sigma^+(&010),K=+1,N,S,I,G,F_1|\nonumber\\
    \times &H_\text{EQM}|^2\Sigma^+(010),K'=-1,N,S,I,G,F_1\rangle\bigg|\nonumber\\
    =\bigg|\frac{2eQq_2}{4\sqrt{6}}&\times\mathcal{G}(K=+1,K'=-1,N,N,S,I,G,G,F_1)\bigg|,
\end{align}
where
\begin{align}
    \mathcal{G}&(K,K',N,N',S,I,G,G',F_1)\nonumber\\
    &\equiv(-1)^{N+G+F_1}\SixJ{G'}{N'}{F_1}{N}{G}{2}\nonumber\\
&\times\sum_{q=\pm2}(-1)^{N-K}\ThreeJ{N}{2}{N'}{-K}{q}{K'}\sqrt{(2N+1)(2N'+1)}\nonumber\\
&\times(-1)^{G'+I+2+S}\sqrt{(2G+1)(2G'+1)}\SixJ{I}{G'}{S}{G}{I}{2}\nonumber\\
&\times\ThreeJ{I}{2}{I}{-I}{0}{I}^{-1}.\label{eq:MatrixElementG}
\end{align}
$\mathcal{G}(K,K',N,N',S,I,G,G',F_1)$ contains information about the alignment of the nuclear spin with the internuclear axis.
Fig.~\ref{fig:G_function} shows the calculated $\mathcal{G}(K,K',N,N',S,I,G,G',F_1)$ with $N'=N$ and $G'=G$ for $^{173}$YbOH ($I=5/2$).
The alignment factor is $\mathcal{O}(1)$ depending on $N$, $G$, $I$, and $F_1$.

\setlength{\tabcolsep}{3.5pt} 
\renewcommand{\arraystretch}{1.2} 
\begin{table*}[ht!]
\centering
\begin{threeparttable}
\caption{Parity doubling without nuclear EQM $\omega_{\mathcal{P};0}$ and with nuclear EQM $\omega_\mathcal{P;\text{EQM}}$. $I$: nuclear spin, $Q$: nuclear EQM, and $q_2$: quadratic electric field gradient parameter. Nuclear EQM values are taken from \cite{EQMTable2021}. Values with ``$\sim$" are estimates based on \cite{OmegaDoublingAndLimit2014}.
The computed values of $\omega_{\mathcal{P};\text{EQM}}$ should be regarded as order-of-magnitude estimates.
$\mathcal{F}(\Omega,\Omega',I,J,J',F_1)$ and $\mathcal{G}(K,K',N,N',S,I,G,G',F_1)$ are $\mathcal{O}(1)$ factors containing the information about nuclear spin alignment with the intermolecular axis, and are defined in Eq.~\eqref{eq:MatrixElementF} and Eq.~\eqref{eq:MatrixElementG}, respectively.}
\begin{tabular}{|c |c |c |c |c |c ||c| c|} 
 \hline
 molecule & $I$ & $Q$($10^{-28}$m$^2$) &state & $q_2\left(\frac{e}{4\pi\epsilon_0a_B^3}\right)$ &$\frac{\omega_{\mathcal{P};0}}{2\pi}$(MHz) &\multicolumn{2}{c|}{$\frac{\omega_\mathcal{P;\text{EQM}}}{2\pi}$ (MHz)} \\
\hhline{========}
   $^{177}$HfF$^+$ &7/2&3.37& $\TDO(J=1)$&0.0535   & 0.74 \cite{cossel2012broadband}  & ~8.6~&\multirow{6}{*}{{\makecell{$\times\mathcal{F}(\Omega=+1,\Omega'=-1,I,J,J'=J,F_1)$\\\text{[}Eq.~\eqref{eq:MatrixElementF}\text{]}}}}\\\cline{1-7}
   $^{179}$HfF$^+$ &9/2&3.79& $\TDO (J=1)$&0.0535& 0.74 \cite{cossel2012broadband}         & ~9.7~&\\\cline{1-7}
   $^{181}$TaN &7/2&3.17& $\TDO (J=1)$&0.0378     &  $\sim1$& ~5.7~&\\\cline{1-7}
   $^{181}$TaO$^+$ &7/2&3.17& $\TDO (J=1)$&0.0135&  $\sim1$& ~2.0~&\\\cline{1-7}
  $^{229}$ThO  &5/2&3.11& $\TDO(J=1)$&0.1537        & 0.36 \cite{ThOHStateCharacterization2011} & ~23~&\\\cline{1-7}
   $^{229}$ThF$^+$ &5/2&3.11& $\TDO (J=1)$&0.0371 & 5.5 \cite{ThF+Spectroscopy2022}  &~5.3~ &\\\hline
   $^{173}$YbOH &5/2&2.80& $^2\Sigma^+ (010)~(N=1)$&0.0005& 12 \cite{YbOHCharacterizing_2023} & ~0.06~&\multirow{3}{*}{\makecell{$\times\mathcal{G}(K=+1,K'=-1,N,N'=N,$\\~~~~~~~~~$S,I,G,G'=G,F_1)$\\\text{[}Eq.~\eqref{eq:MatrixElementG}\text{]}}}\\\cline{1-7}
   $^{175}$LuOH$^+$ &7/2&3.49& $^2\Sigma^+ (010)~(N=1)$&0.0314& $\sim10$ & ~5.3~&\\\cline{1-7}
   $^{223}$RaOH &3/2&1.22& $^2\Sigma^+ (010)~(N=1)$& 0.0010& $\sim10$ & ~0.06~&\\ 
   \hline
\end{tabular}
   \label{table:EQMSummary}
   \end{threeparttable}
\end{table*}

\section{Results and Discussion}
\label{sec:Discussion}
Table~\ref{table:EQMSummary} summarizes the EQM-induced parity doubling for the molecules we analyzed.
The EQM-induced parity doubling $\omegaPEQM$ is listed together with the parity doubling without an EQM, $\omega_{\mathcal{P};0}$.  
The relative sign between them is specific to each molecule and state and we do not discuss it further.
As discussed in Sec.~\ref{sec:OmegaDoublet}, the nuclear EQM modifies parity doubling, and this effect is significant in some molecules.
Using the lab-frame Stark shift for each molecule ($d_\text{lab}=0.5\text{--}1$~MHz/(V/cm) for the listed molecules), one can calculate the required electric field $\mathcal{E}_\mathcal{P}$ [Eq.~\eqref{eq:Ep}].

The table is separated into two categories.  
The first includes molecules with an $\Omega$-doublet science state. 
The largest EQM-induced effect is for $^{229}$ThO with $\omegaPEQM/(2\pi) = 23~\text{MHz}\times\mathcal{F}(\Omega=+1,\Omega'=-1,I,J,J'=J,F_1)$, about two orders of magnitude larger than the parity doubling without EQM $\omega_{\mathcal{P};0}$.
Since $^{229}$ThO is a neutral molecule, even with the larger parity-doubling, it still seems possible to polarize it using indium-tin-oxide-coated electric field plates~\cite{ACME2Result,RamseyBeamMethod1950}.  

For molecular ions ($^{177}$HfF$^+$, $^{179}$HfF$^+$, $^{181}$TaO$^+$, and $^{229}$ThF$^+$), the required electric field could be more demanding.
The state-of-the-art JILA EDM trap can apply a maximum electric field of $\sim100$~V/cm~\cite{JILASystematic2023}.  
Ideally, one would apply a field several times higher than $\mathcal{E}_\mathcal{P}$ to fully polarize the molecules.
In the current ion trap EDM protocol, the $\pi/2$ pulse in the Ramsey sequence also utilizes the lab-frame rotating electric field and thus depends on $\omegaP$~\cite{JILADemonstration2013}.  
Enhanced parity doubling could affect this scheme.
How significantly it will affect depends on the quantum states used in the Ramsey sequence is specific for each molecule and state.

Another concern with larger parity doubling is the increased electric-field-dependent differential $g$-factor.
Due to the mixing induced by the Stark shift, two doublet science states will have slightly different $g$-factors\cite{ACMETechnicalPaper2017,JILASystematic2023}
\begin{equation}
    \delta g \equiv 2\eta|\mathcal{E}_\text{lab}|.
\end{equation}
This differential $g$-factor makes the doublet sensitive to magnetic fields.
For instance, $\eta=-7.9\times10^{-8}/(\text{V/cm})$ in $^{232}$ThO\cite{ACMETechnicalPaper2017} and $\eta=5.7\times10^{-8}/(\text{V/cm})$ in $^{180}$HfF$^+$\cite{JILASystematic2023}.
While the differential $g$-factor is not yet a dominant systematic effect in $^{232}$ThO, it is one of the important systematic effects in $^{180}$HfF$^+$.  
For example, for $^{177}$HfF$^+$ and $^{179}$HfF$^+$, the EQM-induced parity doubling $\omegaPEQM$ is about 10 times larger than the $\omega_{\mathcal{P};0}$, which will require a larger electric field and potentially result in a larger systematic error.

The second category includes molecules that form parity doublets by vibrational angular momentum ($\ell$-doublet).  
As discussed in Sec.~\ref{sec:Calculation}, since the EQM-induced coupling is through the RT effect, the effect is typically smaller than electronic parity doubling---except for $^{175}$LuOH$^+$.  
We found a large RT coupling in $^{175}$LuOH$^+$, generating an EQM-induced parity doubling $\omegaPEQM$ comparable to $\omega_{\mathcal{P};0}$~\cite{LuOH2022Proposal,LuOHMQMEstimate2020}.  
In $^{173}$YbOH and $^{223}$RaOH, the EQM-induced doubling $\omegaPEQM$ is smaller than the $\omega_{\mathcal{P};0}$ and is not dominant.  
In $^{173}$YbOH, the $\ell$-doubling in $\tilde{X}^2\Sigma(010)$ has been measured at 12.5(5)MHz\cite{YuikiMagicField2023}, consistent with 12.0(2)MHz in $^{174}$YbOH\cite{YbOHCharacterizing_2023}.
These are consistent with our calculations indicating negligible EQM contribution.

Some octupole-deformed nuclei, such as $^{225}$Ra and $^{227}$Th, have spin $I=1/2$ and therefore no EQMs.  
They still have an enhanced sensitivity to NSM due to octupole deformation.
The effect from the nuclear magnetic hyperfine coupling on the parity doubling is also estimated in App.~\ref{sec:dipole} and appears to be negligible.
For example, $^{225}$RaOH~\cite{LaserCoolablePoly2017}, $^{225}$RaOCH$_3$ and $^{225}$RaOCH$_3^+$~\cite{RaOCH3Phelan2021} have large nuclear enhancement factors for NSM due to its octupole deformation, making them attractive candidates~\cite{RaOctupoleEstimate2020,RaOctupoleObservation2019,RaFIsotopeShift2021,NeutronRichRaFSpectroscopy2018,RaFNatureSpectroscopy_2024,RaFNature_2020}.  
$^{227}$Th is also considered to have a large octupole deformation~\cite{Th227OctupoleEvidence2021,Th227ReflectionAsymmetry1995,K1/2Th227_2002}, yielding a high sensitivity to NSM.  
$^{227}$ThO and $^{227}$ThF$^+$ have the same $\TDO$ $\Omega$-doublet structure as $^{232}$ThO, $^{180}$HfF$^+$, and $^{232}$ThF$^+$, making them also good candidates.
The listed molecules (but not limited to these) can be controlled with advancing quantum control techniques to search for symmetry-violating physics.

In all molecules, the EQM coupling also induces large splitting in $\Pi$ states.  
Unlike $\TDO$ $\Omega$-doublet and $^2\Sigma^+ (010)$ $\ell$-doublet, this coupling directly induces GHz-scale parity doubling.
These excited states are primarily used as intermediate states for state preparation and state readout in symmetry-violation searches~\cite{ACMETechnicalPaper2017,JILASystematic2023}.
The large shifts can be absorbed into laser detunings, and we do not expect them to impact the measurement protocols significantly.

Finally, we briefly mention several proposals that do not rely on parity doublets:
$^{138}$BaF~\cite{BaFEDMProposal2018}, $^{173}$YbF~\cite{YbFMQM2023}, $^{205}$TlF~\cite{CENTREXPRoposal_2021}, and $^{223}$FrAg~\cite{FrAgRaAgCalculation2021}.
Although the EQM can still perturb the internal structure of these molecules, similar to the case in $\Pi$ states, its effect can be incorporated into the effective Hamiltonian and detuning of laser frequencies.
We have not identified any significant modifications due to the EQM in these systems.
\section{Conclusion}
\label{sec:Conclusions}
We have calculated EQM-induced parity doubling for several molecules proposed for searches of $P$- and $T$-violating physics.
The magnitude of the effect is first estimated from qualitative discussions and then obtained from \textit{ab initio} calculations.
We find that the EQM-induced contribution is significant in some electronic parity doublets (e.g., $\Omega$-doublets) and also non-negligible in some vibrational doublets ($\ell$-doublets).
We have also discussed how the enhanced parity doubling impacts experimental protocols due to the increased requirement for polarizing electric field.

\section*{acknowledgements}
This work is supported by the National Science Foundation under Grant No.\ PHY-2409434 and PHY-2136573. The work at Johns Hopkins University is supported by the National Science Foundation under Grant No.\ PHY-2309253 (L.\ C.).  We thank D. DeMille, J. M. Doyle, M. Verma, Y. Takahashi, C. Zhang, E. A. Cornell, K. B. Ng, Y. Zhou, A. Jadbabaie, N. R. Hutzler, P. Yu, C. Diver, and  A. Hiramoto for fruitful discussions and feedback. X. F. is supported by the Masason Foundation.

\appendix
\section{Nuclear-Magnetic-Moment-induced Parity Doubling}
\label{sec:dipole}
In addition to the EQM-induced coupling, nuclear spin also generates a nuclear spin–electron spin dipole-dipole interaction~\cite{Brown_Carrington_2003}.
\begin{equation}
    H_\text{dip}=-\frac{\sqrt{10}\mu_0g_S\mu_Bg_N\mu_N}{4\pi}T^1(\boldsymbol{S},\boldsymbol{C}^2)\cdot T^1(\boldsymbol{I}),
\end{equation}
where $\mu_0$ is the vacuum permeability, $g_S$ is the electron magnetic moment~\cite{ElectronMagneticMoment_Fan_PRL_2023}, $\mu_B$ is the Bohr magneton, $g_N$ is the $g$-factor of the nucleus, and $\mu_N$ is the nuclear magneton.  
The matrix element for a Hund's case (a) state is \cite{Brown_Carrington_2003}
\begin{align}
    \langle\eta&,S,\Sigma,\Lambda,J,\Omega,I,F_1|T^1(\boldsymbol{S},\boldsymbol{C}^2)\cdot T^1(\boldsymbol{I})|\eta',S,\Sigma',\Lambda',J',\Omega',I,F_1\rangle\nonumber\\
    =&-\sqrt{3}(-1)^{J'+F_1+I}\SixJ{I}{J'}{F_1}{J}{I}{1}\sqrt{I(I+1)(2I+1)}\nonumber\\
    &\times\sum_q(-1)^{J-\Omega}\ThreeJ{J}{1}{J'}{-\Omega}{q}{\Omega'}\sqrt{(2J+1)(2J'+1)}\nonumber\\
    &\times\sum_{q_1,q_2}(-1)^q\ThreeJ{1}{2}{1}{q_1}{q_2}{-q}(-1)^{S-\Sigma}\ThreeJ{S}{1}{S}{-\Sigma}{q_1}{\Sigma'}\nonumber\\
    &\times\sqrt{S(S+1)(2S+1)}\,\langle\eta,\Lambda|\frac{C^2_{q_2}(\theta,\phi)}{r^{3}}|\eta',\Lambda'\rangle,\label{eq:dipolarHarmoinics}
\end{align}
where $C^\ell_q(\theta,\phi)\equiv\sqrt{\frac{4\pi}{2\ell+1}}Y^\ell_q(\theta,\phi)$ is the modified spherical harmonic function\cite{Brown_Carrington_2003}.
Eq.~\eqref{eq:dipolarHarmoinics} induces a coupling with $\Delta\Sigma = \mp1$, $\Delta\Lambda = \pm2$, and $\Delta\Omega = \pm1$  
($q_1 = \pm1$, $q_2 = \mp2$, and $q = \mp1$, respectively).  
Two of these couplings can generate a path that connects $|\Omega=\pm1\rangle$;
\begin{align}
    |\Sigma,\Lambda,\Omega\rangle&=|-1,+2,+1\rangle\nonumber\\
    &\xrightarrow{\text{dip}}|0,0,0\rangle\xrightarrow{\text{dip}}|+1,-2,-1\rangle.\label{eq:EQM_dip}  
\end{align}
Denoting the magnitude of the dipole-induced mixing as $A_\text{dip}$, the induced parity doubling is given by
\begin{equation}
    \omega_{\mathcal{P};\mathrm{dip}}\simeq\frac{A_\mathrm{dip}^2}{\Delta\omega_e}.
\end{equation}
This effect is large when the energy interval $\Delta\omega_e$ between $|\Sigma,\Lambda,\Omega\rangle = |-1,+2,+1\rangle$ and $|0,0,0\rangle$ is small. 
One of the exceptionally small $\Delta\omega_e$ is between the $\TDO$ and $^1\Sigma^+$ states in ThF$^+$, $\Delta\omega_e= 314~\text{cm}^{-1}$.  
Using the same approach as in Sec.~\ref{sec:Abinitio}, we have evaluated $A_\text{dip}$ between the $\TDO$ and $^1\Sigma^+$ states of ThF$^+$ to be 0.001~cm$^{-1}$ (similarly small for other molecules as well).
The induced parity doubling even in this case is less than 1~kHz and is negligible.
\bibliography{refs}

\end{document}